\documentclass[twocolumn,a4paper,reprint,amssymb,aps,prd,groupedaddress,superscriptaddress,nofootinbib]{revtex4-1}
\pdfoutput=1
\usepackage{multirow}
\usepackage{physics}
\usepackage{dcolumn}
\usepackage{bm}
\usepackage{amsmath}
\usepackage{mathtools}
\usepackage{amsfonts}
\usepackage{pbox}
\usepackage{color}
\usepackage[dvipsnames]{xcolor}
\usepackage{tabularx}
\usepackage{subfigure}
\usepackage{graphicx}
\usepackage{psfrag}
\usepackage{lipsum}
\usepackage{enumitem}
\usepackage{changes}
\usepackage{array}
\usepackage{CJK}
\usepackage[english]{babel}

\usepackage[colorlinks = true,
            linkcolor = blue,
            urlcolor  = blue,
            citecolor = blue,
            anchorcolor = blue]{hyperref}
\usepackage{blindtext}
\usepackage[utf8]{inputenc}
\usepackage{graphicx}
\usepackage{placeins}
\usepackage{siunitx}
\usepackage{amssymb}
\usepackage{mathtools}
\usepackage{physics}
\usepackage{tikz}
\usepackage[titletoc]{appendix}
\usepackage{listings}
\usepackage{float}
\usepackage{braket}

\usepackage[capitalize]{cleveref}
\usepackage{soul}
\usepackage{array}
\newcommand{\be}{\begin{equation}}
\newcommand{\ee}{\end{equation}}

\newcommand{\mincir}{\raise
-3.truept\hbox{\rlap{\hbox{$\sim$}}\raise4.truept\hbox{$<$}\ }}
\newcommand{\magcir}{\raise
-3.truept\hbox{\rlap{\hbox{$\sim$}}\raise4.truept\hbox{$>$}\ }}

\usepackage[dvipsnames]{xcolor}
\definecolor{darkgreen}{rgb}{0., 0.65, 0.1}
\hypersetup{colorlinks,linkcolor={blue},citecolor={red},urlcolor={violet}}

\begin{document}
\title{New late-time constraints on $f(R)$ gravity}

\author{Suresh Kumar}
\email{suresh.math@igu.ac.in}
\affiliation{Department of Mathematics, Indira Gandhi University, Meerpur, Haryana 122502, India}

\author{Rafael C. Nunes}
\email{rafadcnunes@gmail.com}
\affiliation{Instituto de F\'{i}sica, Universidade Federal do Rio Grande do Sul, 91501-970 Porto Alegre RS, Brazil}
\affiliation{Divis\~ao de Astrof\'isica, Instituto Nacional de Pesquisas Espaciais, Avenida dos Astronautas 1758, S\~ao Jos\'e dos Campos, 12227-010, SP, Brazil}

\author{Supriya Pan}
\email{supriya.maths@presiuniv.ac.in}
\affiliation{Department of Mathematics, Presidency University, 86/1 College Street, Kolkata 700073, India}
\affiliation{Institute of Systems Science, Durban University of Technology, PO Box 1334, Durban 4000, Republic of South Africa}

\author{Priya Yadav}
\email{priya.math.rs@igu.ac.in}
\affiliation{Department of Mathematics, Indira Gandhi University, Meerpur, Haryana 122502, India}

\begin{abstract}
Modification of general relativity (GR) inspired by theories like $f(R)$ gravity is among the most popular ones to explain the late-time acceleration of the Universe as an alternative to the $\Lambda$CDM model. In this work, we use the state-of-the-art BAO+BBN data and the most recent Type Ia supernovae (SNe Ia) sample namely PantheonPlus, including the Cepheid host distances and covariance from SH0ES samples, to robustly constrain the $f(R)$ gravity framework via two of the most popular $f(R)$ models in literature, namely, the Hu-Sawicki and Starobinsky models.  Additionally, we consider how the time variation of the Newton's gravitational constant affects the supernovae distance modulus relation. We find a minor evidence for $f(R)$ gravity under the Hu-Sawicki dynamics from BAO+BBN and BAO+BBN+uncalibrated supernovae joint analysis, but the inclusion of Cepheid host distances, makes the model compatible with GR. Further, we notice tendency of this model to relax the $H_0$ tension. In general, in all the analyses carried out in this study with the late time probes, we find both the $f(R)$ models to be consistent with GR at 95\% CL.  
\end{abstract}

\maketitle
\section{Introduction}
\label{sec:intro}
Astronomical data are precious for modern cosmology. 
From the detection of the cosmic microwave background anisotropy to the late-time dynamics of the Universe, we have witnessed the crucial role played by the astronomical data. For instance, 
the dynamics of our Universe at its late time got abruptly changed since 1998 from the observations of Type Ia supernovae (SNe Ia)  which  first reported  one of the trailblazing results in modern cosmology \textemdash{} the accelerating expansion of our Universe~\cite{SupernovaSearchTeam:1998fmf,SupernovaCosmologyProject:1998vns}. 
This late-time accelerating expansion  demands that a revision of the standard cosmology is essential and we need to invoke some exotic type of fluids into the gravitational equations. 
This can be done effectively by two distinct ways, either one can modify the matter sector of the Universe without touching the gravitational sector described by the Einstein's General Relativity (GR) which leads to various Dark Energy (DE) models
\cite{Peebles:2002gy,Copeland:2006wr,Sahni:2006pa,Bamba:2012cp,Li:2012dt,Motta:2021hvl}, or the Einstein's GR can be modified in various ways, known as modified gravity (MG)  theories   \cite{Nojiri:2006ri,Sotiriou:2008rp,DeFelice:2010aj,Clifton:2011jh,Capozziello:2011et,Koyama:2015vza,Cai:2015emx,Nojiri:2017ncd,Ferreira:2019xrr,Bahamonde:2021gfp}. Following both the approaches, 
over the last several years, a cluster of DE and MG models have been tested with the available astronomical data (see Refs. \cite{Copeland:2006wr,Bamba:2012cp,Li:2012dt,Nojiri:2006ri,Sotiriou:2008rp,DeFelice:2010aj,Clifton:2011jh,Capozziello:2011et,Koyama:2015vza,Cai:2015emx,Nojiri:2017ncd,Ferreira:2019xrr,Bahamonde:2021gfp} and the references therein).

Among the existing DE and MG models,  the $\Lambda$-Cold Dark Matter ($\Lambda$CDM) cosmological model, where $\Lambda >0$ acts as a DE candidate in the context of GR, is an excellent cosmological model that fits to a large span of astronomical datasets. Nevertheless, $\Lambda$CDM cosmology faces many theoretical and observational challenges. Recent observations of $\Lambda$CDM-based Planck 2018~\cite{Aghanim:2018eyx} and the SH0ES (Supernovae and $H_0$ for the Equation of State of dark energy) collaboration~\cite{Riess:2021jrx,Riess:2022mme}  suggest that the Hubble constant $H_0$ from Planck $\Lambda$CDM is at more than $5 \sigma$ tension with the SH0ES measurement  \cite{Riess:2021jrx,Riess:2022mme}. In addition, measurements of the parameter $S_8$  ($= \sigma_8\sqrt{\Omega_{\rm m}/0.3}$; $\sigma_8$ is amplitude of the matter power spectrum and $\Omega_{\rm m}$ is the matter density parameter at present time)  estimated by the Planck 2018 \cite{Aghanim:2018eyx}, weak lensing experiments~\cite{Heymans:2020gsg,KiDS:2020ghu,DES:2021vln,DES:2022ygi} and Redshift-Space Distortions  measurements \cite{Kazantzidis:2018rnb,Skara:2019usd,Nunes:2021ipq} are in tension  at more than $3 \sigma$. These suggest that a revision of the $\Lambda$CDM cosmology is needed to agree with the observational evidences. As a consequence, several alternative proposals to the $\Lambda$CDM cosmology appeared to explain such observational discrepancies \cite{DiValentino:2016hlg,Yang:2018qmz,Vagnozzi:2019ezj,Visinelli:2019qqu,Alestas:2020mvb,DiValentino:2020naf,Yang:2021flj,Yang:2021hxg,Kumar:2021eev,Chudaykin:2022rnl,Ballardini:2023mzm} (see the recent reviews in this direction \cite{DiValentino:2021izs,Perivolaropoulos:2021jda,Schoneberg:2021qvd,Kamionkowski:2022pkx}). However, despite many new and appealing cosmological models,  it has been observed that simultaneous solution to both the tensions are quite difficult to obtain \cite{Abdalla:2022yfr}. Thus, understanding  the nature of the cosmological tensions and their solutions demands further attention through new observational probes and cosmological models. 

In this article we focus on one of the viable alternatives to the $\Lambda$CDM cosmology \textemdash{} the  modified gravity theory, where in particular, we consider the most natural modification to the Einstein's GR, namely, the $f(R)$ gravity which has been greatly investigated considering both the theoretical and observational perspectives \cite{Bessa:2021lnr,Leizerovich:2021ksf,Farrugia:2021zwx,Pan:2021tpk,Negrelli:2020yps,Capozziello:2018aba,Akarsu:2018ykf,Lazkoz:2018aqk,Nunes:2016drj,Nunes:2016drj}, as well as  to assuage the current cosmological tensions \cite{DAgostino:2020dhv,Wang:2020dsc,Odintsov:2020qzd}. However, unlike in the past works, our approach in this work significantly differs in the treatment of its observational analysis that enters through the physics of SNe Ia, 
 and such a difference is caused due to the consideration of new scalar degree(s) of freedom beyond GR which results in a time dependent Newton's gravitational constant $G$. 
 Such a varying $G$ may induce a redshift ($z$-) dependent effect on the peak luminosity of SNe Ia from the mass of the white dwarf progenitors \cite{Amendola:1999vu,Gaztanaga:2001fh,Wright:2017rsu} and this may result in changes in the cosmological constraints of the modified gravity models. The revision in the evolution of intrinsic luminosity of SNe Ia due to variation of $G$ has been considered to constrain several cosmological models \cite{Perivolaropoulos:2021bds,Alestas:2021luu,Sapone:2020wwz,Kumar:2022nvf,Ballardini:2021eox}. 

This means that for precise understanding of the cosmology of modified gravity theories, the impact of modified gravity theories on the astrophysics of SNe Ia should be considered,  and through the estimation of the cosmological parameters using the modified formalism, such impact can be decoded. Following this, the key aim of this article is to employ the above modifications to constrain the $f(R)$ gravity models, and study the resulting implications mainly in light of the $H_0$ tension. To test this hypothesis, we use for the first time the Pantheon+ sample to constrain the free parameters of the $f(R)$ gravity models. In addition to these perspectives, we also consider for the first time in this work how the state-of-the-art assumptions on BAO+BBN joint analysis can constrain the behavior of the $f(R)$ gravity models at late times.

The article is organized as follows. In Sec. \ref{sec-fR-description+model}, we provide a brief introduction to the cosmology of $f(R)$ gravity and introduce two well known models that we investigate in this article. In Sec. 
 \ref{sec-data}, we describe the observational data-sets and our methodology to constrain the baseline of the proposed  $f(R)$ gravity models. In Sec. \ref{sec-fR-results},  we describe the observational constraints on the $f(R)$ models and discuss our main results. Finally, we describe our conclusions and perspectives in Sec. \ref{sec-summary}.

\section{$f(R)$ gravity and cosmology}
\label{sec-fR-description+model}

The gravitational action of $f(R)$ gravity, in Jordan frame, is given by  
\begin{equation}
\mathcal{S} = \frac{1}{16\pi G}\; \int d^4 x \sqrt{-g}\,\, f(R) + \mathcal{S}_{\rm m}+ \mathcal{S}_{\rm r}\,,
\label{action}
\end{equation}
where $R$ denotes the Ricci scalar and $G$ is the Newton's gravitational constant. Additionally, eqn. (\ref{action}) includes  
the actions for the matter sector ($\mathcal{S}_{\rm m}$) and the radiation sector ($\mathcal{S}_{\rm r}$).  We assume that there is no interaction at the non-gravitational level between matter sector and the radiation sector, that means both these sectors are independently conserved. 
Now varying the action (\ref{action})
with respect to the metric $g_{\mu\nu}$, we obtain the gravitational equations
\begin{eqnarray}
F G_{\mu\nu}
= -\frac{1}{2} g_{\mu \nu} \left( F R - f (R) \right)
+ \nabla_{\mu}\nabla_{\nu}F -g_{\mu \nu} \Box F  \nonumber\\ +
8\pi G\,  \left[T^{(\rm m)}_{\mu \nu} +T^{(\rm r)}_{\mu \nu}\right]
\,,
\label{gravitational-eqns}
\end{eqnarray}
where  $G_{\mu\nu}= R_{\mu\nu}-\left(1/2\right)g_{\mu\nu}R$ stands for 
the Einstein tensor; ${\nabla}_{\mu}$  is the covariant derivative,
$\Box \equiv g^{\mu \nu} {\nabla}_{\mu} {\nabla}_{\nu}$; 
$F = F(R) \equiv f_{,R}= d f(R)/dR$ (similarly by $f_{, RR}$ we shall mean $d^2 f(R)/dR^2$); $ T^{(\rm m)}_{\mu \nu} $ and $ T^{(\rm r)}_{\mu 
\nu} $ respectively denote  the  energy-momentum tensor for the matter sector and the radiation 
sector.  Note that for $f (R) = R$ in eqn. (\ref{action}), one recovers the Einstein-Hilbert action for General Relativity. Now we proceed towards the cosmological evolution in the context of $f(R)$ gravity theory. As usual, we start with the homogeneous and isotropic background of our Universe which is well described by the Friedmann-Lema\^{i}tre-Robertson-Walker (FLRW) line element

\begin{eqnarray}
\text{d}s^2=-\text{d}t^2+a^2(t)\left[\frac{\text{d}r^2}{1-kr^2}+r^2(\text{d}\theta^2+\sin^2\theta
\,\text{d}\phi^2)\right], \label{flrw}
\end{eqnarray}
where $(t, r, \theta, \phi)$ are the co-moving coordinates; $a(t)$ describes the expansion scale factor of the Universe and $k$ corresponds to the spatial geometry of the Universe where $k =0$, $+1$ and $-1$, respectively denote a spatially flat, closed and open Universe.    
Now for the spatially flat FLRW line element ($k =0$),  eqn. (\ref{gravitational-eqns}) leads to 

\begin{eqnarray}
3FH^2
=
8\pi G  \left(\rho_\text{m}+\rho_\text{r}\right) +\frac{1}{2} \left( F R - f (R) \right)
-3H\dot{F}\,,
\label{FE-1} \\
-2F \dot{H}
= 8\pi G  \left( \rho_\text{m} + p_\text{m} +\rho_\text{r} + p_\text{r} \right)
+\ddot{F}-H\dot{F} \,,
\label{FE-2}
\end{eqnarray}
where an overhead dot denotes the derivative with respect to the cosmic time $t$; $H = \dot{a} (t)/a(t)$ is the Hubble parameter; ($\rho_\text{m}$, $p_\text{m}$), $(\rho_\text{r}, p_\text{r})$  denote the (energy density, pressure) of the matter sector and the radiation sector respectively. Note that in the spatially flat FLRW Universe, the Ricci scalar $R$ takes the form $R = 6 (2H^2+\dot{H} )$. It might be interesting to note that the gravitational equations (\ref{FE-1}) and (\ref{FE-2}) can also be expressed as $3H^2=8\pi G  \left(\rho_m +\rho_r+\rho_{\rm eff}\right)$ and $2\dot{H}= - 8\pi G  \left( \rho_m + p_m + \rho_r+ p_r+ \rho_{\rm eff} + p_{\rm eff}\right)$ respectively, where through $\rho_{\rm eff}$, $p_{\rm eff}$, defined as 
\begin{eqnarray}
\rho_{\rm eff}\equiv
\frac{1}{8\pi G } \left[
\frac{1}{2} \left( FR - f \right) -3H \dot{F} +3\left(1-F\right)H^2
\right], \label{rho-eff}\\
p_{\rm eff} \equiv \frac{1}{8\pi G }\Bigg[ -\frac{1}{2} \left( FR - f \right)
-\left(1-F\right)\left(2\dot{H}+3H^2\right) \nonumber\\ + \ddot{F}+2H \dot{F} 
\Bigg], \label{p-eff}
\end{eqnarray}
one introduces an effective dark energy scenario from the modifications of the gravitational sector where the effective equation-of-state parameter $w_{\rm eff} \equiv p_{\rm eff}/\rho_{\rm eff}$ reads
\begin{eqnarray}
 w_{\rm eff} = -1 - \frac{H \dot{F}+ 2 \dot{H}-2F\dot{H}- \ddot{F}}{\frac{1}{2} \left( FR - f \right) -3H \dot{F} +3\left(1-F\right)H^2},   
\end{eqnarray}
which represents a deviation from the cosmological constant $w_{\Lambda} = -1$ induced by the gravitational modifications. Thus, for any modified $f(R)$ gravity model, one can calculate the effective equation-of-state and estimate how far the model is deviating from the cosmological constant.  
One can notice that $\rho_{\rm eff}$ and $p_{\rm eff}$ defined in eqns. (\ref{rho-eff}), (\ref{p-eff}) 
satisfy the usual conservation equation  $\dot{\rho}_{\rm eff}+3H(\rho_{\rm eff}+p_{\rm eff})=0$. As the matter and radiation sectors enjoy independent conservation, therefore, their conservation equations can be expressed as 
\begin{eqnarray}
&&\dot{\rho}_\text{m}+3H(1+w_\text{m})\rho_\text{m}=0, \label{balance-matter}\\
&&\dot{\rho}_\text{r}+3H (1+w_\text{r})\rho_\text{r}=0, \label{balance-radiation}
\end{eqnarray}
 where $w_\text{m} = p_\text{m}/\rho_\text{m}$ and $w_\text{r} = p_\text{r}/\rho_\text{r}$ are respectively the equation of state parameters of the matter sector and the radiation sector. We assume the standard cases where $w_\text{m} =0$ (i.e., pressure-less matter) and $w_\text{r} = 1/3$. Therefore, from eqns. (\ref{balance-matter}) and (\ref{balance-radiation}), one can derive that $\rho_\text{m}  \propto a^{-3}$ and $\rho_\text{r}  \propto a^{-4}$, respectively.

Now, for a given $f(R)$ model,  using the gravitational equations (\ref{FE-1}) and (\ref{FE-2}) together with the conservation equations for the matter and radiation sectors, in principle, one can determine the cosmological dynamics. However, an arbitrary $f(R)$ model may suffer from a number of cosmological problems, e.g. the matter instability \cite{Faraoni:2006sy}, instability at the level of perturbations \cite{Bean:2006up}, 
absence of matter dominated era \cite{Amendola:2006kh}, inability to satisfy the local gravity constraints \cite{Chiba:2006jp}, dark energy oscillations \cite{Nojiri:2006ww} etc. Thus, in order to construct viable $f(R)$ models, one needs to impose the following conditions \cite{DeFelice:2010aj,Amendola:2006we}: 

\begin{figure*} [hbt!]
    \centering
    \includegraphics[width=8cm]{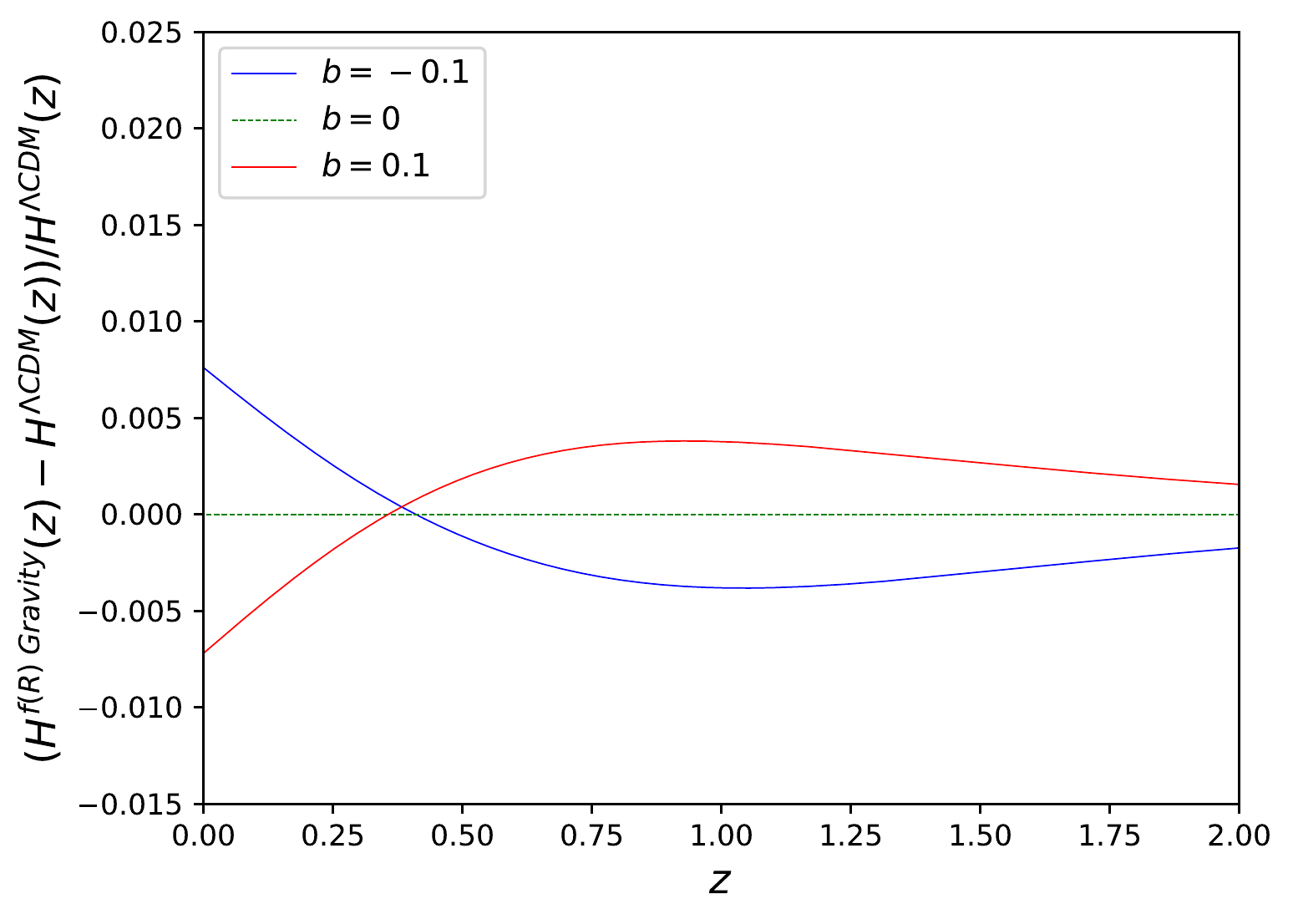} \,\,\,\,\,\,
    \includegraphics[width=8cm]{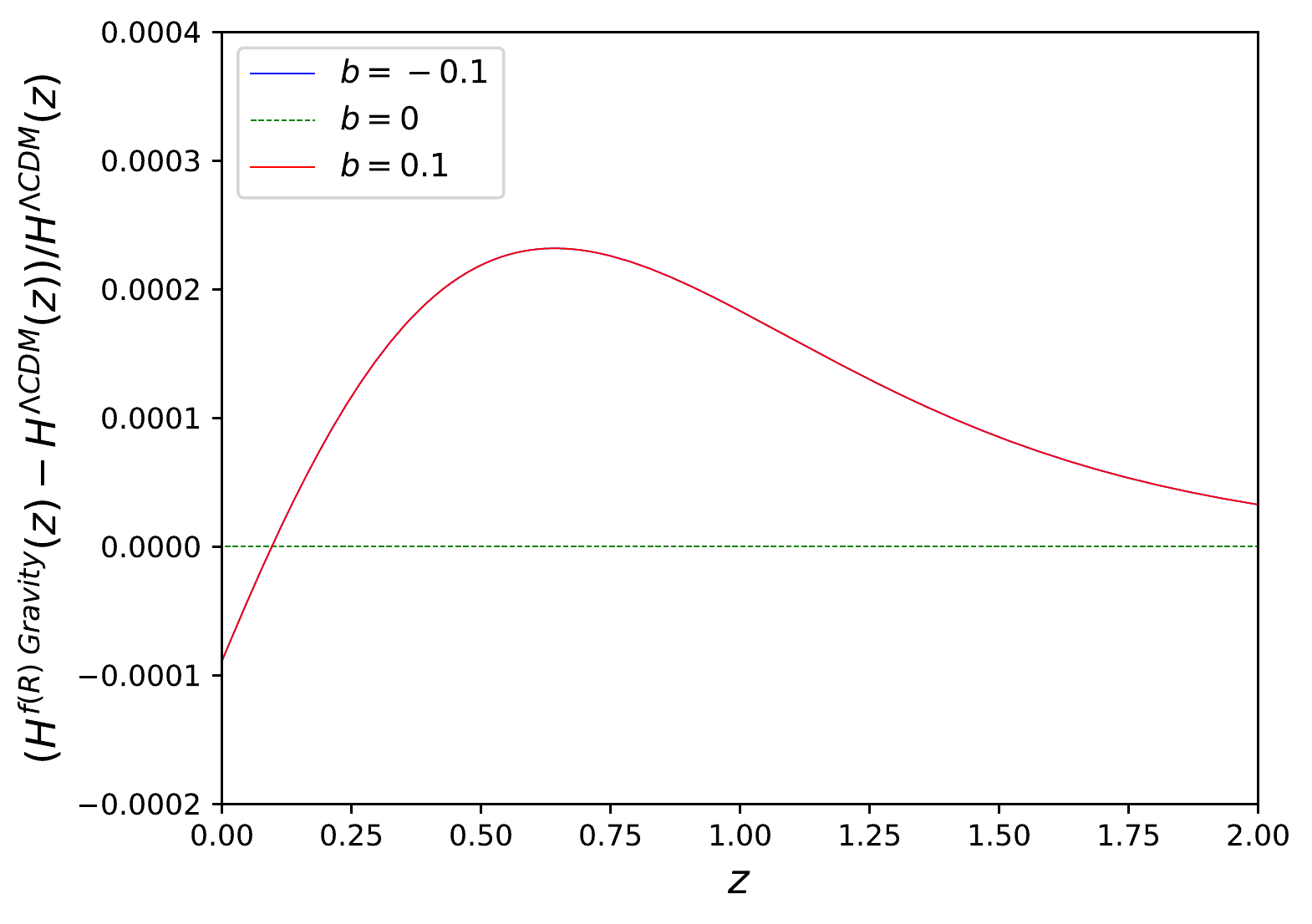}
\caption{Left panel: Relative difference in the expansion rate of the Universe, $\Delta H(z)= \left(H^ {f(R)\, {\rm Gravity}}(z)/H^{\Lambda {\rm CDM}}(z) \right)-1$, for the Hu-Sawicki $f(R)$ model  considering several values of $b$. Right panel: Same as in left panel, but for the Starobinsky $f(R)$ model. }
   \label{fig:deltaH}
\end{figure*}

\begin{eqnarray}
f_{, R} > 0\; \text{and}\;  f_{,RR} > 0\;\mbox{for}~R \geq R_0~ (>0),
\end{eqnarray}
where $R_0$ is the present value of $R$. The condition $f_{,R} > 0$ ensures that there are no ghosts and $f_{,RR} >0$ ensures the avoidance of tachyonic instability \cite{DeFelice:2010aj}. 
Moreover, from the observational perspectives, a viable $f(R)$ model reproducing the matter dominated era, satisfying the local gravity constraints plus to be consistent with the equivalence principle, should behave like 
\begin{eqnarray}
f(R)\rightarrow R-2\Lambda, \; \text{for}\ \ 
R\geq R_0,
\end{eqnarray}
where $\Lambda $ is a constant and to depict a late-time stable  de 
Sitter solution \cite{DeFelice:2010aj}, the $f(R)$ model also needs  to satisfy 
\begin{eqnarray}
0< \left(\frac{R f_{,RR}}{f_{,R}}\right)_{r} <1 \ \ \text{at}\ \ 
r=-\frac{Rf_{,R}}{f}=-2.
\end{eqnarray}
Combining all these conditions altogether, the viable $f (R)$ models up to 
two parameters can be recast as 
\begin{eqnarray}
f(R)=R-2\Lambda y(R,b),
\label{fRyRb}
\end{eqnarray}
where the function $y(R,b)$ gives an idea about the deviation of the underlying $f (R)$ model from GR in which $b$ is a free parameter. 
In the following, we consider two viable $f(R)$ models, namely the Hu-Sawicki $f(R)$ model \cite{Hu:2007nk} and the Starobinsky $f(R)$ model  \cite{Starobinsky:2007hu}.

\begin{enumerate}
 
\item The Hu-Sawicki $f(R)$ model reads as \cite{Hu:2007nk}

\begin{eqnarray}
f(R)= 
R -
\frac{c_1 R_{\mathrm{HS}} \left(R/R_{\mathrm{HS}}\right)^p}{c_2
\left(R/R_{\mathrm{HS}}\right)^p + 1},
\label{HSfR}
\end{eqnarray}
where $c_1$, $c_2$, $R_{\mathrm{HS}}$ and $p~(>0)$ are the free parameters of the model. 
One can rewrite eqn. (\ref{HSfR}) to the form of eqn. (\ref{fRyRb}) where $y(R, b)$ adopts the following expression  \cite{Bamba:2012qi,Basilakos:2013nfa,Nunes:2016drj}
\begin{equation}
y(R, b) = 1- \frac{1}{1+ \Bigl(\frac{R}{\Lambda b} \Bigr)^p},
\end{equation}
  in which  $c_1R_{\mathrm{HS}}/c_2 = 2 \Lambda$ and $2c_2^{1-1/p}/c_1 = b $.  In this article we consider $p =1$ as usual, since this parameter is completely statistically degenerate. Now, one may notice that for $b \rightarrow 0$ (i.e., for $c_1 \rightarrow \infty$) and $R_{\mathrm{HS}}  \rightarrow 0$ with $c_1 R_{\mathrm{HS}} \rightarrow 2 \Lambda c_2$,  the Hu-Sawicki $f(R)$
model can also approximate the $\Lambda$CDM cosmology, i.e., $f(R) \rightarrow R - 2 \Lambda$.

\item The Starobinsky dark energy $f(R)$ model is given by \cite{Starobinsky:2007hu}.

\begin{eqnarray}
f(R)=
R -
\lambda R_{\mathrm{S}} \left[ 1 -
\left(1+\frac{R^2}{R_{\mathrm{S}}^2} \right)^{-n}
\right],
\label{StarfR}
\end{eqnarray}
where $\lambda~(>0)$, $R_{\mathrm{S}}$  and $n~(>0)$ are the free parameters of this model.  In a similar fashion,  one can rewrite 
eqn. (\ref{StarfR}) to the form of eqn. (\ref{fRyRb}) where $y(R, b)$ takes the form \cite{Basilakos:2013nfa,Nunes:2016drj}
\begin{equation}
y(R, b) = 1- \frac{1}{\Bigl[1+ \left(\frac{R}{\Lambda \, b} \right)^2\Bigr]^n},
\end{equation}
in which $\Lambda=  \lambda R_{\mathrm{S}}/2$ and $b= 2/\lambda$.  Throughout the article we have considered $n =1$ as usual, since this parameter is completely statistically degenerate.  Now, one may notice that for $b \rightarrow 0$ (i.e., for $\lambda \rightarrow \infty$) and $R_{\mathrm{S}}  \rightarrow 0$ with $\lambda R_{\mathrm{S}} \rightarrow 2 \Lambda$,  the Starobinsky $f(R)$
model can also approximate the $\Lambda$CDM cosmology, i.e., $f(R) \rightarrow R - 2 \Lambda$. Note that this model does not have the same functional structure as the famous Starobinsky inflation model \cite{Starobinsky:1980te}, but represents a parametric form that allows to generate an accelerated expansion at late times, where for $R >> R_s$ we have $f(R) = R - 2\Lambda$, where the high curvature value of the effective cosmological constant is $\Lambda = \lambda R_s/2$. 
 
\end{enumerate}

Thus, one can see that the free parameter $b$ quantifies the deviation from GR ($b =0$). In Appendix A, we show the scalar potential for both models in the Einstein frame. Now, in order to understand the evolution of the Universe 
for the proposed $f(R)$ models, one needs to trace the expansion rate of the Universe.  We use the same methodology as in Refs. \cite{Basilakos:2013nfa,Sultana:2022qzn} to derive the expansion rate of the Universe, i.e., the $H(z)$ function, for the proposed $f(R)$ models.  Fig. \ref{fig:deltaH} shows the theoretical prediction for the  expansion rate of the Universe at late times for both the models under consideration in this work taking reasonable and different values of $b$. We quantify the difference  
from the $\Lambda$CDM model by using the fixed values of $H_0$ and $\Omega_{\rm m}$ to their canonical values from CMB observations \cite{Planck:2018vyg}, i.e., $H_0=67.4$ km/s/Mpc and $\Omega_{\rm m}=0.31$. For the Hu-Sawicki model (see the left panel of Fig. \ref{fig:deltaH}), we note that the expansion of the Universe is very sensitive to $b$, irrespective of the positive or negative values as clearly depicted here for $b \in [-0.1, 0.1]$. Specifically, for $b > 0$ and $z > 0.38$, the expansion rate of the Universe within the Hu-Sawicki $f(R)$ model 
is greater than the $\Lambda$CDM model while for $z < 0.38$, we notice the inverse scenario. For $b < 0$, the dynamics is opposite to the previous case assuming $b > 0$. On the other hand, for the Starobinsky $f(R)$ model (see the right panel of Fig. \ref{fig:deltaH}),  we see that irrespective of the positive and negative values of $b$ within $[-0.1, 0.1]$, the behaviour in the expansion rate within this $f(R)$ gravity model remains similar. In fact, the expansion rate of the Universe within this $f(R)$ model is unresponsive to variations in the sign of parameter $b$ with respect to the $\Lambda$CDM model.  Overall, we find that for $z > 0.12$, the expansion rate of the Universe is higher than the $\Lambda$CDM model while for $z < 0.12$, the expansion rate of the Universe is lower than the $\Lambda$CDM model.  In summary, one can see that the Hu-Sawicki and the Starobinsky $f(R)$ models are quantitatively not the same at late times.

\section{Data and Methodology}
\label{sec-data}

In order to derive constraints on the model baseline, we use the following datasets.

\begin{itemize}
\item \textbf{BAO}: Baryon Acoustic Oscillation (BAO) data consist of isotropic BAO measurements of $D_V(z)/r_d$, where $D_V(z)$ and $r_d$ stand for spherically averaged volume distance, and sound horizon at baryon drag respectively and anisotropic BAO measurements of $D_M(z)/r_d$ and $D_H(z)/r_d$ (with $D_M(z)$ the comoving angular diameter distance and $D_H(z)=c/H(z)$ the Hubble distance) from the final measurements of the SDSS collaboration that cover eight distinct redshift intervals, acquired and ameliorated over the past 20 years \cite{eBOSS:2020yzd}. All the above mentioned BAO-only measurements are compiled in Table 3 of Ref. \cite{eBOSS:2020yzd}. We assume that the uncertainties are Gaussian approximations to the likelihoods for each tracer ignoring the correlations between measurements as suggested in the SDSS collaboration
paper \cite{eBOSS:2020yzd}.

\item \textbf{BBN}: The  Big Bang Nucleosynthesis (BBN) are considered with the state-of-the-art assumptions, which consist of measurements of the primordial abundances of helium, $Y_P$, from~\cite{Aver:2015iza}, and the deuterium measurement, $y_{DP} = 10^5 n_D/n_H$, obtained in~\cite{Cooke:2017cwo}. This BBN likelihood is sensitive to the  physical baryon density $\omega_b \equiv \Omega_bh^2$ and the effective number of neutrino species $N_{\rm eff}$ constraints. In the present work, we fix $N_{\rm eff} = 3.046$ .

\item \textbf{Type Ia supernovae and Cepheid}: Type Ia supernovae (SNe Ia) have generally been an important astrophysical tools in establishing the standard cosmological model. SNe Ia distance moduli measurements constrain the uncalibrated luminosity distance $H_0d_L(z)$, or in other words the slope of the late-time expansion rate, which as a result constrains the matter density parameter $\Omega_{\rm m}$. For a supernova at redshift $z$, the theoretical apparent magnitude $m_B$ is given by

\begin{eqnarray}
\label{distance_modulus}
m_B = 5 \log_{10} \left[ \frac{d_L(z)}{1 Mpc} \right] + 25 + M_B,
\end{eqnarray}
where $M_B$ is the absolute magnitude. The distance modulus reads as $\mu(z) = m_B - M_B$.
The calibrated SNe Ia absolute magnitude $M_B$ is in general assumed to be truly a constant, i.e., the parameter $M_B$ should  be independent of the redshift. It has been argued that a possible variation of the absolute magnitude $M_B$ and equivalently of the absolute luminosity as $L \sim 10^{-2M_B/5}$, could be due to a variation in the value of Newton’s gravitational constant $G$ \cite{Gaztanaga:2001fh,Wright:2017rsu}. This is due to the fact that the absolute luminosity is proportional to the Chandrasekhar mass as $L \sim M_{\rm Chandra}$,  which depends on $G$ as $L \sim G^{-3/2}$. Therefore, any modification of gravity will generate an  effective gravitational constant in the form of $G_{\rm eff}$ that will induce a natural correction to the distance modulus. The presence of a varying effective gravitational constant leads to rewrite eq. (\ref{distance_modulus}) as
\begin{eqnarray}
\label{mb_fr}
 &&\mu_{\rm th} = m_B - M_B\nonumber\\ && \;\; ~~~= 5 \log_{10} d_L(z) + 25 + \frac{15}{4} \log_{10} \frac{G_{\rm eff}(z)}{G}.
\end{eqnarray}

Taking the quasi-static approximation and the modified Poisson equation, it is well known that in $f(R)$ gravity context, we have \cite{Tsujikawa:2007gd}

\begin{equation}
\frac{G_{\rm eff}(z)}{G} = \frac{1}{f_R} \left(\frac{1+4 k^2 m/a^2}{1+3k^2 m/a^2} \right),
\end{equation}
where $m = f_{RR}/f_R$ and $G_{\rm eff} (z)$ is the effective gravitational constant in the $f(R)$ gravity framework. 

\begin{figure}
\includegraphics[width=9cm]{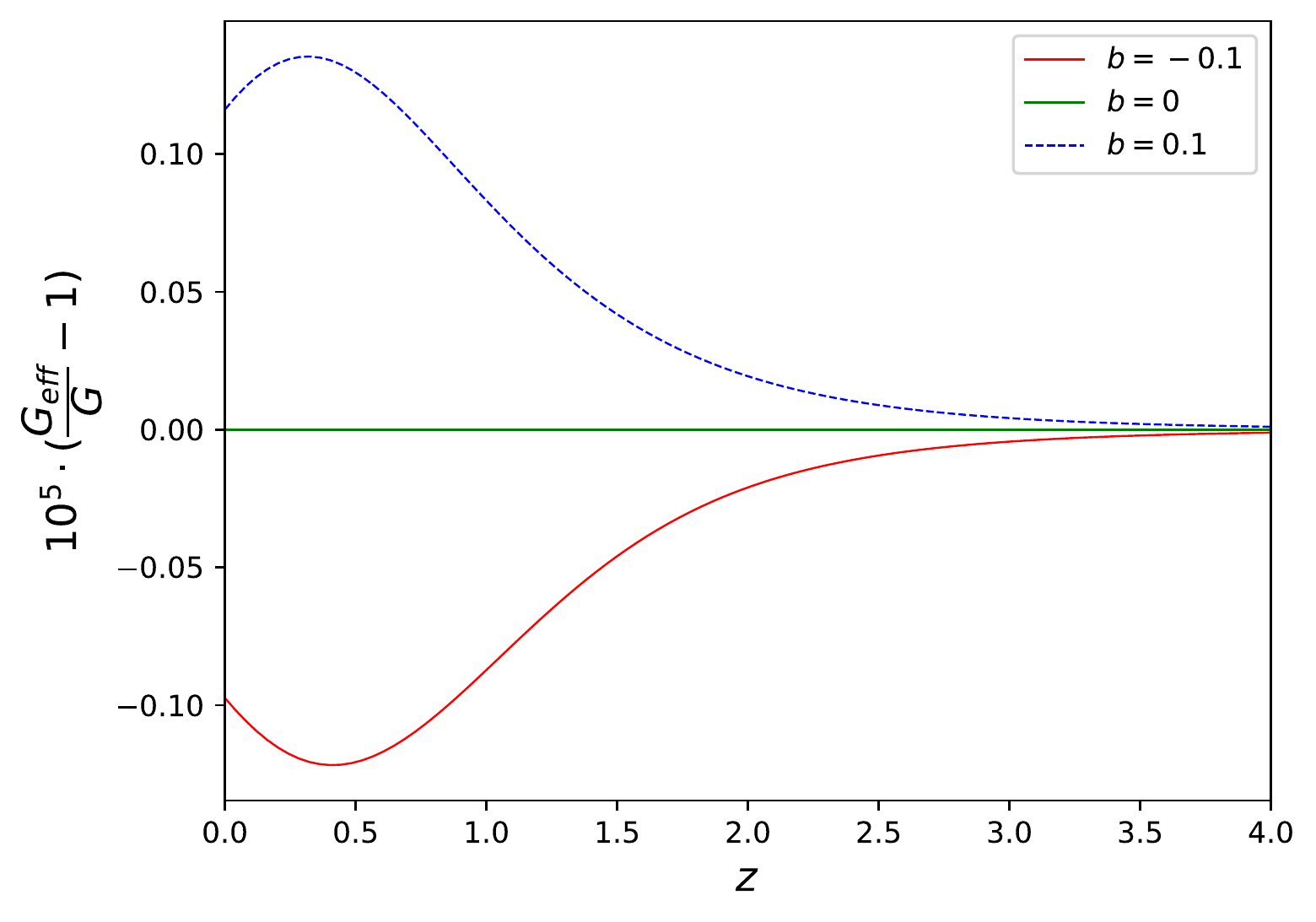} 
\caption{ The effective gravitational coupling $G_{\rm eff}(z)/G$
as a function of redshift $z$ for different and reasonable values
of $b$ for the Hu-Sawicki parameterization within the framework adopted in this work. }
\label{Geff_HS}
 \end{figure}

Note that eq. (\ref{mb_fr}) reduces to GR when $f(R) = R - 2\Lambda$, i.e., the $\Lambda$CDM model. We follow Refs. \cite{Basilakos:2013nfa,Nesseris:2017vor} and set $k = 0.1$ h/Mpc, which is necessary as now the Newton's gravitational constant depends on the scale $k$ as well. In Fig. \ref{Geff_HS} we show the relative difference on the effective gravitational coupling $G_{\rm eff} (z)$ for the reasonable values of the quantifying parameter $b$ under the perspective of the Hu-Sawicki  model. We can note that the effective gravitational constant has effects of the order of $10^{-4}$ within the adopted fix range. On the other hand, for the Starobinsky model, we note that $G_{\rm eff}$ has corrections of the order of $10^{-11}$, and hence this can 
be considered to have negligible effects.

In \cite{delaCruz-Dombriz:2008ium}, the authors discuss the validity of some parameterizations obtained in the quasi-static approximation. In Ref. \cite{Orjuela-Quintana:2023zjm}, it is demonstrated  that the 
quasi-static approximation for the Hu-Sawicki model is valid for the scale 0.01 h/Mpc $< k < 0.2$ h/Mpc, which is in accordance with the values adopted in our analyses. The validity of the quasi-static approximation using the N-body simulation has been examined in \cite{Bose:2014zba}. The quasi-static regime for the Starobinsky model was investigated in \cite{Gannouji_2009}. Therefore, we can consider these two most popular scenarios in literature as stable theories within these perspectives. Note that, in principle, 
the effective gravitational constant is a function of scale as well, and this could have a significant effect on the CMB predictions  on 
very large scale, where the late-time integrated Sachs Wolfe effect has strong contributions. Then the $f(R)$ gravity could contribute with extra corrections due to that fact besides only the
change on the expansion rate  of the Universe and metric potential corrections on very large scale at CMB level. Also, the adopted scale $k$ values coincide with the cosmic scales where the oscillation effects 
are predominant on the matter power spectrum, which could, in principle, also affect the model-dependent predictions measurements of BAO. Both points are not yet addressed in the literature, and we hope to check it out in future communications.

We use the SNe Ia distance moduli measurements from the
Pantheon+ sample \cite{Brout:2022vxf}, which consists of 1701 light curves
of 1550 distinct SNe Ia ranging in the redshift interval $z \in [0.001, 2.26]$, publicly available at \url{https://pantheonplussh0es.github.io/}. We refer to this dataset as \texttt{PantheonPlus}. We also consider the
SH0ES Cepheid host distance anchors, which facilitate constraints on both $M_B$ and $H_0$. When utilizing SH0ES Cepheid host distances, the SNe Ia distance residuals are modified following the relationship eq.(14) of Ref.  \cite{Brout:2022vxf}. We refer to this dataset as \texttt{PantheonPlus\&SH0ES}. 

Thus, it is possible that the modification on the  distance moduli  induced from the $f(R)$ gravity framework may carry useful information about the dynamics of these scenarios. 

\end{itemize}

     In our analyses, we allow the parameters $\omega_b$, $\omega_{\rm cdm}$ (physical cold dark matter density), $H_0$,  $b$ and $M_B$ (in the analyses with PantheonPlus data) with wide ranges of flat priors. Note that the extra correction in eq. (\ref{mb_fr}) can be interpreted as a new and time dependent absolute magnitude in the form of $M_B(z) = M_B + 15/4 \;\rm log_{10}(G_{\rm {eff}}$ $(z)/\rm G)$, where $M_B$ is a constant parameter. Several cosmological tests have recently been done to test the robustness of the constancy of the Supernova absolute magnitude $M_B$ \cite{Benisty:2022psx,Sapone:2020wwz,Tutusaus:2017ibk,Martinelli:2019krf,DiValentino:2020evt}, where no clear statistical evidence proves that $M_B$ can be time-dependent. \\
     
     For $G_{\rm eff}/G$ = 1, we recover the default case. As default from a statistical point of view, we treat $M_B$ as a free and nuisance parameter in all analyses. We ran \texttt{CLASS+MontePython} code~\cite{classI,Blas_2011,Audren:2012wb,Brinckmann:2018cvx} using Metropolis-Hastings mode to derive constraints on the cosmological parameters for the $f(R)$ gravity models defined in Sec. \ref{sec-fR-description+model} using several combinations of the datasets. All of our runs reached a Gelman-Rubin convergence criterion of $R - 1 < 10^{-2}$.  Further, we use \texttt{MCEvidence}\footnote{\href{https://github.com/yabebalFantaye/MCEvidence}{github.com/yabebalFantaye/MCEvidence}} algorithm to compute the Bayesian evidence and perform a model comparison through the Jeffreys’ scale \cite{vazquez2012bayesian}. For model comparison, we use the log-Bayesian evidence for each of the models relative to the standard $\Lambda$CDM model, i.e., $\Delta \rm ln \mathcal{Z} = \rm ln \mathcal{Z}_{\Lambda CDM} -  \rm ln \mathcal{Z}_{f(R) gravity}$. For the interpretation of the results, we refer to the revised Jeffrey's scale and accordingly  the evidence is inconclusive if $0 \leq |\Delta \rm ln \mathcal{Z}|  < 1$, weak if $1 \leq | \Delta \rm ln \mathcal{Z}|  < 2.5$, moderate if $2.5 \leq |\Delta \rm ln \mathcal{Z}|  < 5$, strong if $5 \leq | \Delta \rm ln \mathcal{Z}|  < 10$, and very strong if $| \Delta \rm ln \mathcal{Z} | \geq 10$ \cite{Kass:1995loi,Trotta:2008qt}. 
In what follows, we discuss the main results of our analyses.

\section{Main results and discussions}
\label{sec-fR-results}

\begin{figure*}
    \centering
    \includegraphics[width=8.9cm]{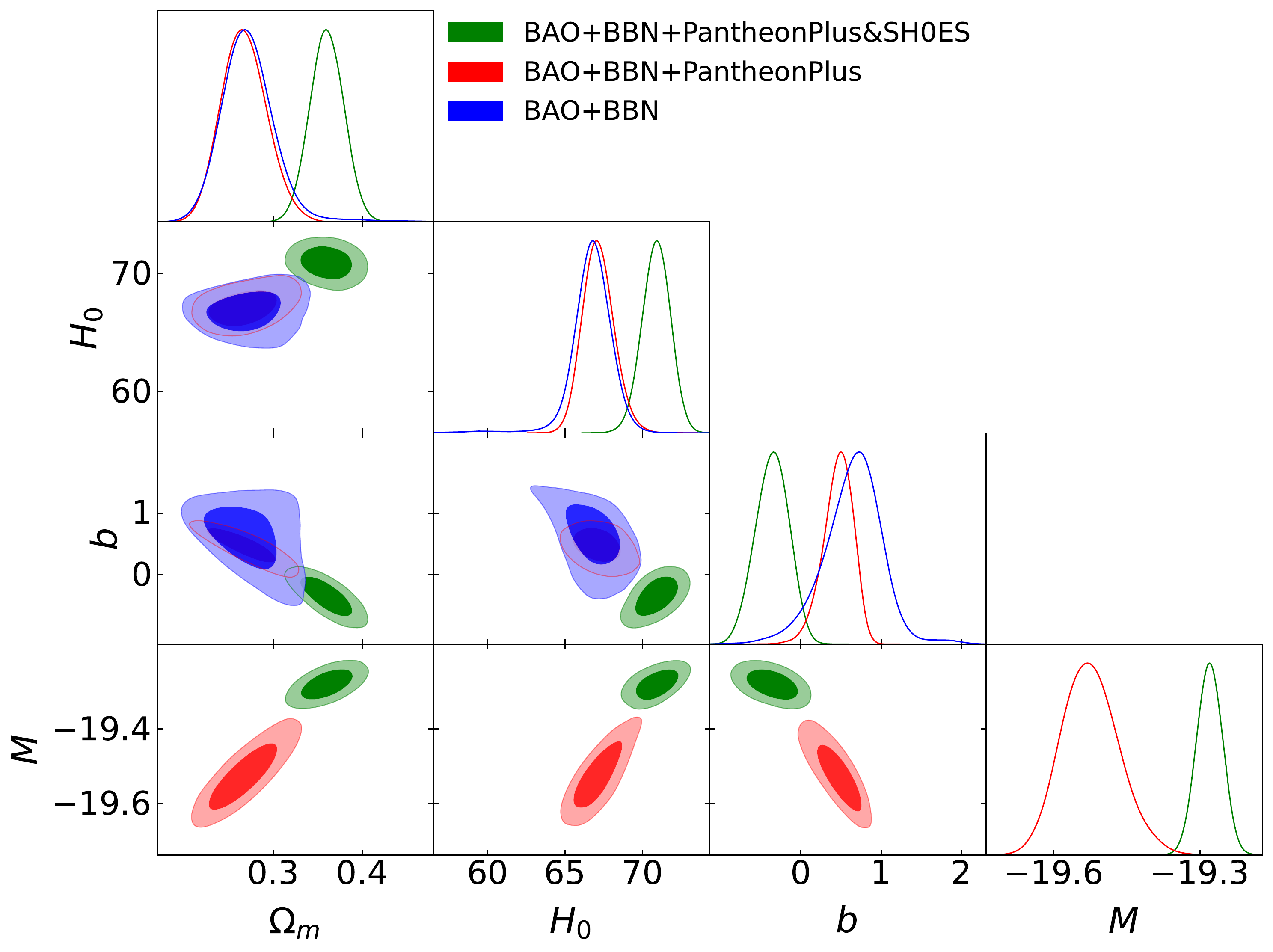}
    \includegraphics[width=8.9cm]{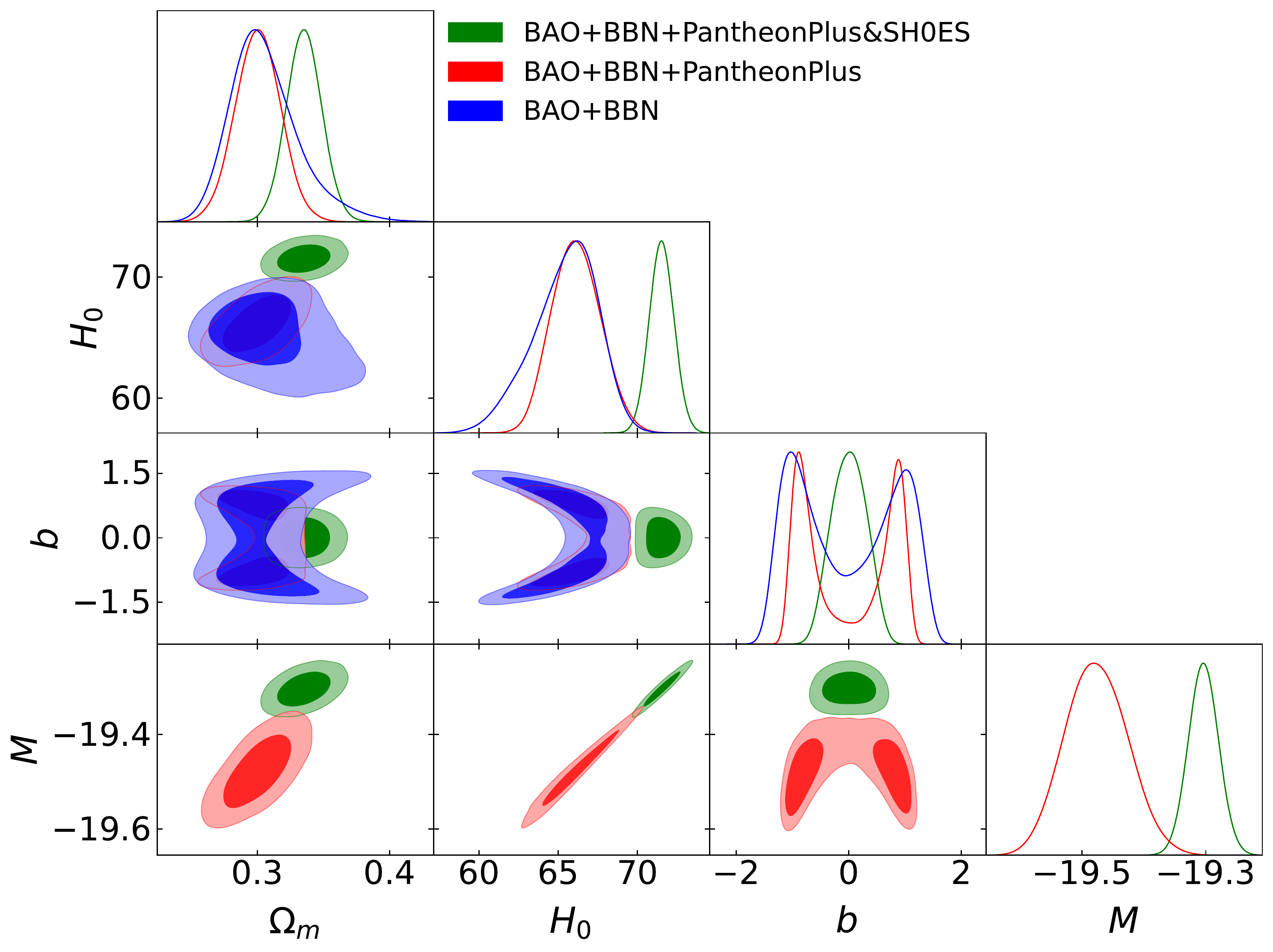}
    \caption{One-dimensional posterior distributions and two-dimensional marginalized confidence regions (68\% CL and 95\% CL) for $b$, $\Omega_{\rm m}$ and $H_0$  obtained from the BAO+BBN, BAO+BBN+PantheonPlus and BAO+BBN+PantheonPlus\&SH0ES for the Hu-Sawicki model (left panel) and Starobinsky model (right panel). The parameter $H_0$ is in units of km/s/Mpc.}
   \label{ps_HS}
\end{figure*}

In Table \ref{tab:all}, we report the summary of the statistical analyses considering the Hu-Sawicki and Starobinsky $f(R)$ models obtained from various observational datasets, namely, BAO+BBN, BAO+BBN+PantheonPlus, BAO+BBN+PantheonPlus\&SH0ES and PantheonPlus\&SH0ES. In addition, we also show the constraints on the $\Lambda$CDM model using the same datasets in Table \ref{tab:all} in order to compare the $\Lambda$CDM results with the Hu-Sawicki and Starobinsky $f(R)$ models. Moreover, in Fig. \ref{ps_HS},  we display the parametric space at 68\% CL and 95\% CL for the Hu-Sawicki (left panel) and Starobinsky (right panel) $f(R)$ models. 

\begin{table*}[hbt!]
\caption{Constraints at 68\% CL  on  some selected parameters of the Hu-Sawicki, Starobinsky and $\Lambda$CDM models obtained from BAO+BBN, BAO+BBN+PantheonPlus, BAO+BBN+PantheonPlus\&SH0ES and PantheonPlus\&SH0ES data. Note that here $\Delta \rm ln \mathcal{Z}$ = $\rm ln \mathcal{Z}_{\Lambda CDM}$ - $\rm ln \mathcal{Z}_{f(R) gravity.}$}
\label{tableI}
\begin{tabular} { |c| c| c| c| c|l|l|  }   
 \hline

Data

&   BAO+BBN  & BAO+BBN+PantheonPlus &   BAO+BBN+PantheonPlus\&SH0ES  & PantheonPlus\&SH0ES \\

 \hline
 Model &  Hu-Sawicki     & Hu-Sawicki  &  Hu-Sawicki     & Hu-Sawicki        \\
 &  \textcolor{blue}{Starobinsky}     & \textcolor{blue}{Starobinsky}  &  \textcolor{blue}{Starobinsky}     & \textcolor{blue}{Starobinsky}       \\
 &  \textcolor{magenta}{$\Lambda$CDM}     & \textcolor{magenta}{$\Lambda$CDM}  &  \textcolor{magenta}{$\Lambda$CDM}     & \textcolor{magenta}{$\Lambda$CDM}       \\
\hline

$b$ &  $0.64^{+0.38}_{-0.29}  $  & $0.46^{+0.21}_{-0.15}            $ &$-0.36\pm 0.21    $ & $-0.003\pm 0.029   $  \\

&   \textcolor{blue}{$-0.02\pm 0.91 $}  & \textcolor{blue}{$-0.02\pm 0.80             $} & \textcolor{blue}{$0.01\pm 0.31      $} &  \textcolor{blue}{$-0.037^{+0.051}_{-0.044}    $} 
\\

& \textcolor{magenta}{0}    &  \textcolor{magenta}{0}
& \textcolor{magenta}{0}    &  \textcolor{magenta}{0}
\\

\hline
$H_0\,[{\rm km}/{\rm s}/{\rm Mpc}]$&  $66.7^{+1.2}_{-1.0}           $ & $67.16^{+0.92}_{-1.1}              $ &$70.89\pm 0.92           $ &  $73.76\pm 0.83              $ 
\\
 
&   \textcolor{blue}{$65.4^{+2.4}_{-1.6} $}  & \textcolor{blue}{$66.1 \pm 1.5             $} & \textcolor{blue}{$71.53\pm 0.77     $} &  \textcolor{blue}{$ 74.0^{+1.0}_{-0.85} $} 
\\

& \textcolor{magenta}{$67.5^{+1.1}_{-1.2}$}    &  \textcolor{magenta}{$68.3^{+1.0}_{-0.91}$}
& \textcolor{magenta}{$71.47\pm 0.68$}    &  \textcolor{magenta}{$73.74\pm 0.98 $}
\\

\hline

$\Omega_{\rm m}$&  $  0.273^{+0.024}_{-0.031}         $ & $  0.268^{+0.024}_{-0.027}            $ &$0.360\pm 0.019          $ &  $   0.334\pm 0.017           $ 
\\
 
&   \textcolor{blue}{$0.306^{+0.019}_{-0.029} $}  & \textcolor{blue}{$0.300 \pm 0.017            $} & \textcolor{blue}{$0.335\pm 0.013  $} &  \textcolor{blue}{$ 0.332\pm 0.019  $} 
\\

& \textcolor{magenta}{$0.297^{+0.017}_{-0.020} $}    &  \textcolor{magenta}{$0.319 \pm 0.013 $}
& \textcolor{magenta}{$0.336\pm 0.013$}    &  \textcolor{magenta}{$ 0.333\pm 0.018$}
\\[1ex]
\hline

$M_B$&  $  -        $ & $  -19.525^{+0.054}_{-0.063}            $ &$-19.280\pm 0.027          $ &  $  -19.241\pm 0.023           $ 
\\
 
&   \textcolor{blue}{$- $}  & \textcolor{blue}{$-19.477 \pm 0.050          $} & \textcolor{blue}{$-19.304\pm 0.024 $} &  \textcolor{blue}{$ -19.235^{+0.029}_{-0.026} $} 
\\

& \textcolor{magenta}{$-$}    &  \textcolor{magenta}{$-19.409^{+0.035}_{-0.031} $}
& \textcolor{magenta}{$-19.307\pm 0.022$}    &  \textcolor{magenta}{$ -19.242\pm 0.028 $}
\\
\hline

 $\Delta \rm ln \mathcal{Z} $&   $-1.62$ & $-1.84$   &  $-0.84$    & $2.33$ 
 \\
 
 &   \textcolor{blue}{$-1.99$}  & \textcolor{blue}{$-2.28$} & \textcolor{blue}{$0.13$} &  \textcolor{blue}{$2.04$} 
\\

& \textcolor{magenta}{$0$}    &  \textcolor{magenta}{$0$}
 & \textcolor{magenta}{$0$}    &  \textcolor{magenta}{$ 0$}
 \\
\hline

\end{tabular}\label{tab:all}
\end{table*}

\begin{table*}[hbt!]
\caption{Constraints at 68\% CL  on  some selected parameters of the Hu-Sawicki and Starobinsky models obtained from  BAO+BBN+PantheonPlus\&SH0ES and PantheonPlus\&SH0ES data without the extra corrections on the distance moduli in eq. (\ref{mb_fr}).}

\begin{tabular} { |c| c| c| c| c|l|l|  }   
 \hline

Data

&     BAO+BBN+PantheonPlus\&SH0ES  & PantheonPlus\&SH0ES \\

 \hline
 Model  &  Hu-Sawicki     & Hu-Sawicki        \\
 &  \textcolor{blue}{Starobinsky}     & \textcolor{blue}{Starobinsky}        \\
\hline

$b$ &  $-0.36^{+0.39}_{-0.42}$  & $-0.09^{+0.19}_{-0.12}$  \\

&    \textcolor{blue}{$0.00\pm 0.31     $} &  \textcolor{blue}{$0.02\pm 0.77      $} 
\\

\hline
$H_0\,[{\rm km}/{\rm s}/{\rm Mpc}]$&  $70.9^{+1.8}_{-1.8}            $ &  $73.5^{+2.1}_{-2.1}                $ 
\\
 
&  \textcolor{blue}{$71.50\pm 0.78     $} &  \textcolor{blue}{$74.10^{+0.97}_{-1.3} $} 
\\

\hline

$\Omega_{\rm m}$&  $0.361^{+0.038}_{-0.037}       $ & $0.347^{+0.042}_{-0.039}         $  
\\
 
&   \textcolor{blue}{$0.335\pm 0.013   $}  & \textcolor{blue}{$0.300^{+0.047}_{-0.022}            $} 
\\

\hline

$M_B$&  $-19.280^{+0.053}_{-0.052}       $ &$-19.236^{+0.060}_{-0.059}            $ 
\\
 
&   \textcolor{blue}{$-19.304\pm 0.024 $}  & \textcolor{blue}{$-19.240\pm 0.028          $} 
\\

\hline

\hline

\end{tabular}\label{tab:alll}
\end{table*}

\begin{figure*}
     \centering
     \includegraphics[width=8cm]{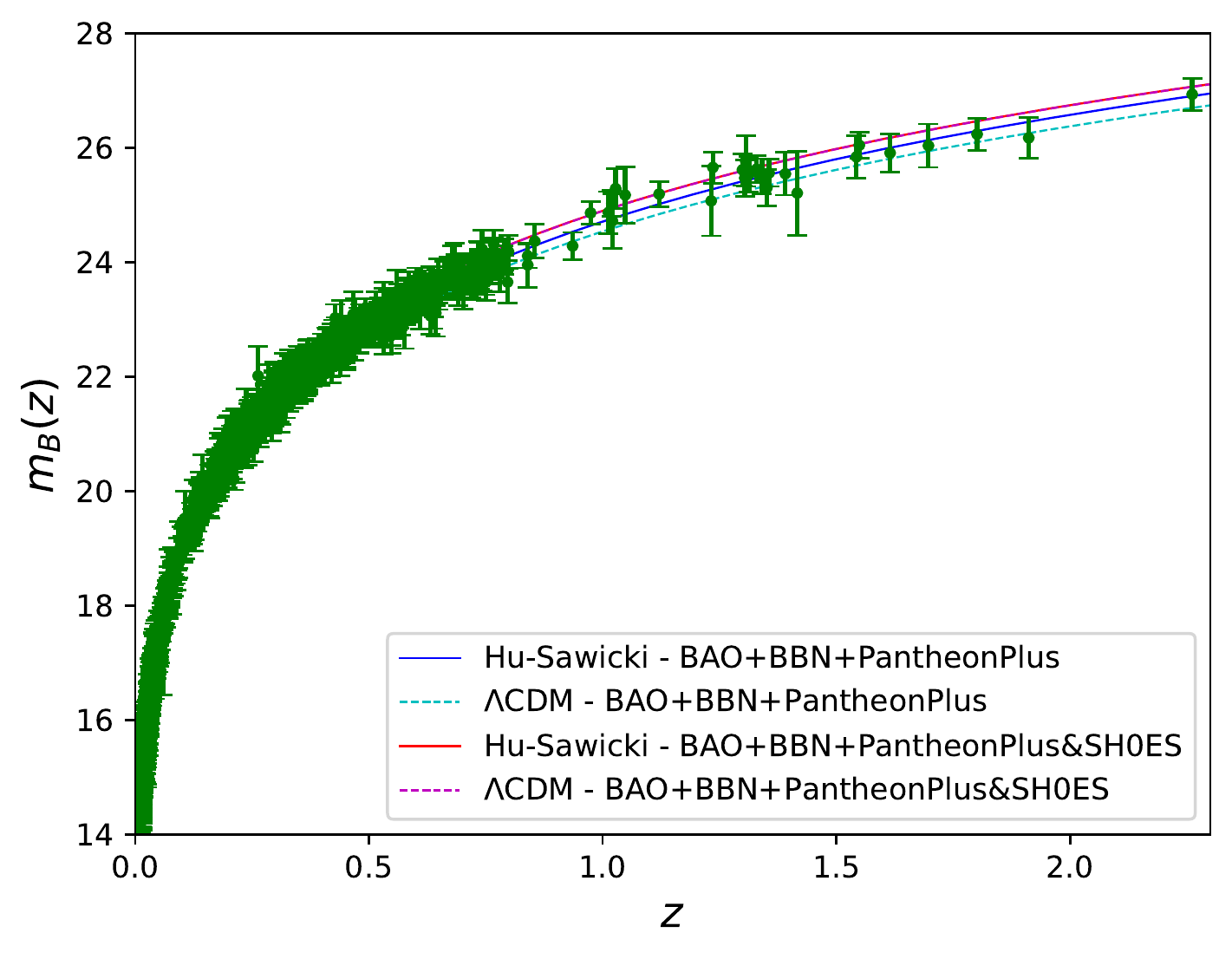} \,\,\,\,\,\,
     \includegraphics[width=8cm]{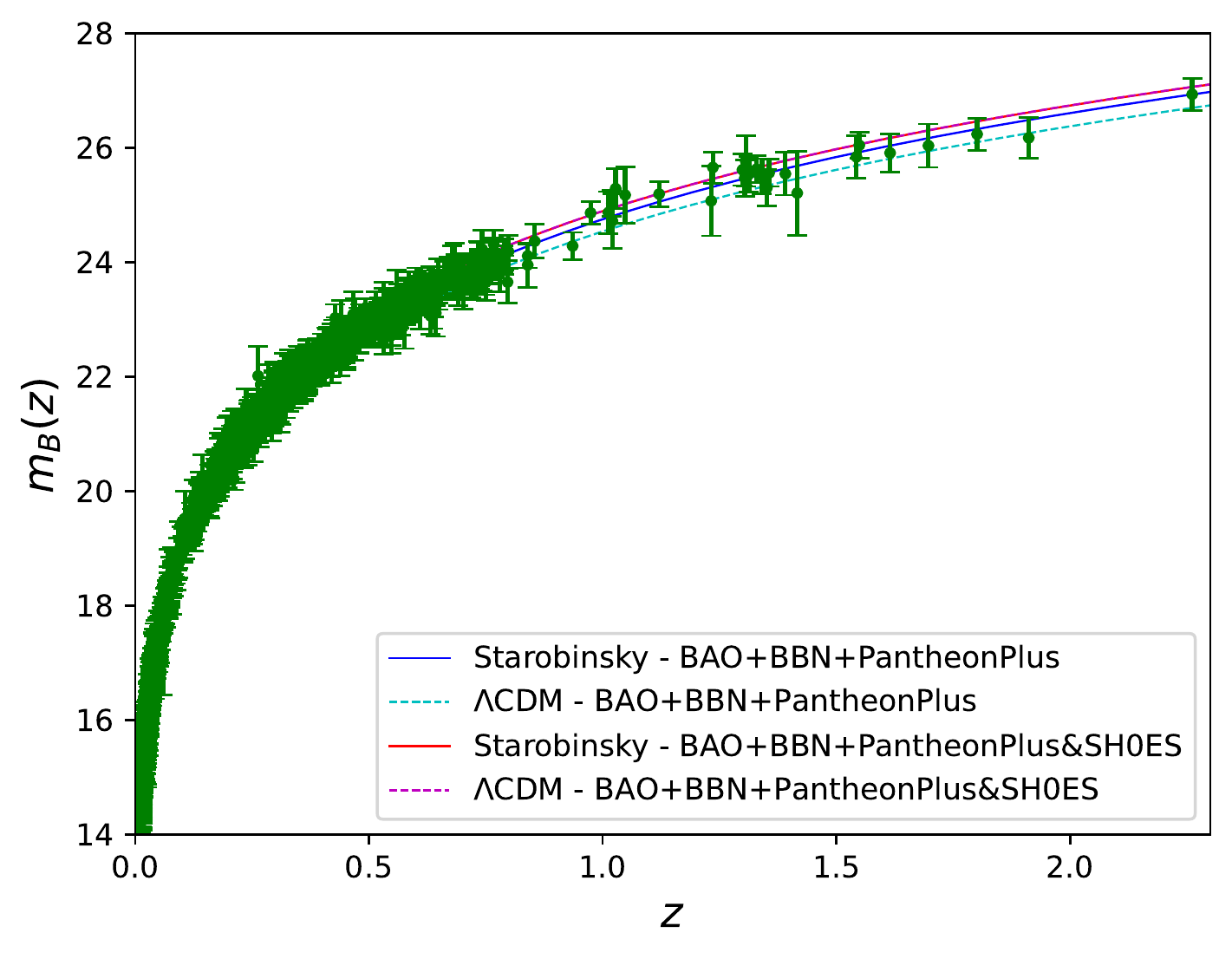}
     \caption{This figure shows the magnitude-redshift relation of the PantheonPlus sample in the range $0 < z < 2.3$ for the best fit values from BAO+BBN+PantheonPlus and BAO+BBN+PantheonPlus\&SH0ES analysis summarized in Table \ref{tab:all} for Hu-Sawicki model (left panel) and Starobinsky model (right panel). The $\Lambda$CDM best-fit prediction is also shown in both panels.} 
    \label{mu_the_ob}
\end{figure*}

The combined dataset BAO+BBN probes the background  history of the model independently of both CMB and supernovae data.
As the $H_0$ tension directly invites a straight conflict between the CMB and the local distance ladder measurements, thus, it will be interesting to find new routes to estimate the Hubble constant. The joint analysis BAO+BBN has been proved to be a competitive cosmological test \cite{Schoneberg:2019wmt,Cuceu:2019for,Schoneberg:2022ggi} 
which can provide accurate confidence limits on the baseline parameters of the models. Thus, we choose BAO+BBN to be our minimum data set.  As well known, the constraints on $H_0$ from BAO+BBN data in the $\Lambda$CDM context fully agree with the CMB data. When applying BAO+BBN in the context of $f(R)$ gravity, we notice this behavior, being $H_0$ compatible with low values obtained in the CMB measurements, and at 1.6\% and 3\% accuracy from the Hu-Sawicki and Starobinsky models, respectively. For the parameter that quantifies the deviation from GR, i.e., the parameter $b$, we find different results in the two different $f(R)$ gravity models.  For the Hu-Sawicki model, we obtain $b > 0$ at more than 68\% CL for BAO+BBN ($b = 0.64^{+0.38}_{-0.29}$ at 68\% CL).   On the other hand, for the Starobinsky $f(R)$ model, we  find that $b$  is compatible to zero within 68\% CL for BAO+BBN,  that means, no deviation from GR is suggested within this $f(R)$ model for BAO+BBN. This is not unexpected because the models have different dynamical behaviors at late times, as previously discussed (see Fig. \ref{fig:deltaH}). We further noticed that the free parameter $b$ in the Starobinsky model is not much sensitive to its sign change  with results being largely symmetric under the change of the sign on the parameter $b$. Thus, the posterior tends to be bimodal based on the prior adopted in our analysis. The addition of SH0ES Cepheid host distances tends only to smooth out the bimodal effect. Because of this bimodality, the estimation of parameter $b$ turns out to be compatible with the null hypothesis, i.e, $b = 0$. Since there are no reasons to impose some hard prior like $b > 0$ and/or $b < 0$, we choose to leave this parameter free within a large prior range.

Now, we move on considering the addition of SNe Ia and Cepheid host distance measurements from the SH0ES team, while considering the additional corrections on the distance moduli, i.e., eq. (\ref{mb_fr}) due to the $f(R)$ gravity model being one of the main motivations of this work. As argued in Refs. \cite{Efstathiou:2021ocp,Camarena:2021jlr,Nunes:2021zzi}, the tension on $H_0$ should be replaced as the tension on the supernova absolute magnitude $M_B$, as the estimate of $H_0$ from SH0ES collaboration comes directly from the estimate of $M_B$. So, in our analysis we first consider  the uncalibrated supernovae sample, i.e., the PantheonPlus. When analyzing with BAO+BBN+PantheonPlus, we find that $b > 0$ at 95\% CL ($b=0.46^{+0.35}_{-0.38}$) for the Hu-Sawicki $f(R)$ model, while for the Starobinksy $f(R)$ model, the null hypothesis $b =0$ is fully compatible within 68\% CL. The constraints on $H_0$ from BAO+BBN+PantheonPlus for both the Hu-Sawicki and Starobinksy $f(R)$ models are fully compatible with the estimates of $H_0$ obtained from BAO+BBN.

Now, following Ref. \cite{Brout:2022vxf} we consider the inclusion of the Cepheid-host distances measurements in direct combination with the SNe Ia sample, i.e., the full dataset PantheonPlus\&SH0ES. We notice that the parameter $b$ in both cases becomes fully compatible with GR, i.e., $b = 0$. Thus, the inclusion of the Cepheid host distances and the full covariance matrix from SH0ES samples, makes the dynamics of the $f(R)$ models similar to $\Lambda$CDM. On the other hand, $H_0$ and $\Omega_{\rm m}$ get larger values compared to the previous analyses without the inclusion of SH0ES measurement. Moreover, it is possible to quantify the level of tension between two estimates $H_{0,i}$ and $H_{0,j}$ of $H_0$ by means of the simple 1-dimensional tension metric, which can be constructed as 
\begin{equation}
T_{H_0} \equiv \frac{|H_{0,i}-H_{0,j}|}{\sqrt{\sigma^2_{H_{0,i}}+\sigma^2_{H_{0,j}}}}\,,
\label{eq:1D_estimator}
\end{equation}
measured in equivalent Gaussian standard deviations. In particular, we find that for the Hu-Sawicki $f(R)$ model (Starobinsky $f(R)$ model), the $H_0$ value from BAO+BBN+PantheonPlus\&SH0ES is at 2.8$\sigma$ (3.2$\sigma$) and 2.9$\sigma$ (2.9$\sigma$) tensions with the $H_0$ values from the BAO+BBN+PantheonPlus and BAO+BBN  analyses, respectively. 

We also consider the PantheonPlus\&SH0ES data without external probes. Considering the fact that this sample is at more than 2$\sigma$ tension with BAO+BBN, we analyze their effects separately. As also shown in Ref. \cite{Brout:2022vxf}, for the flat $\Lambda$CDM, the joint analysis from PantheonPlus\&SH0ES tends to generate high values of $H_0$ (see our results in Table \ref{tab:all}). 
When analyzing PantheonPlus\&SH0ES for the $f(R)$ gravity model, we noticed the same behavior, i.e., $H_0$ gets high values compatible with local measurements. That is, without external probes, $H_0$ inferred from PantheonPlus\&SH0ES sample for non-standard models of type $f(R)$ gravity, $H_0$ is constrained to high values. We note that Ref. \cite{Dhawan_2020} pointed out a possible insensitivity of the local $H_0$ constraint from the Cepheid distance ladder in some model beyond the $\Lambda$CDM cosmology. In this sense, 
and based on the present results, we can conclude that the same is valid for scenarios like $f(R)$ gravity.

It is well known that BAO+SNe Ia
joint analysis prefers low values of $H_0$ compatible with CMB observations, and on the other hand, this joint analysis is in tension with SNe Ia+Cepheid sample analysis (see general discussions introduced in \cite{Efstathiou:2021ocp}). Thus,  from the analyses presented here for BAO+BBN+PantheonPlus and PantheonPlus\&SH0ES, we find that the $f(R)$ gravity does not significantly change the local distance ladder value of $H_0$, and therefore these models are not able to solve the $H_0$ tension in the light of the late time probes.
 In the analyses of the Hu-Sawicki model with the BAO+BBN and BAO+BBN+PantheonPlus data, we notice negative correlation between $H_0$ and $b$ (see left panel of Fig. \ref{ps_HS}). So smaller values of $b$ correspond to the larger values of $H_0$, while we notice opposite scenario in the analysis with BAO+BBN+PantheonPlus\&SH0ES data. On the other hand, in all the analyses of the Starobinsky $f(R)$ model,  the $b$ parameter is insensitive to $H_0$.  Similar and equivalent conclusions can be drawn from the point of view of the $M_B$ estimation. 

It has been shown in the literature that a phantom dark energy model or scenarios that provide an effective phantom behavior at late times, may be the simplest solutions to the Hubble tension problem  \cite{DiValentino:2016hlg,DiValentino:2020naf,Vagnozzi:2019ezj,Yang:2021hxg,Yang:2021flj,Visinelli:2019qqu,Yang:2018qmz,Chudaykin:2022rnl,Ballardini:2023mzm,Kumar:2021eev}. Also it can be seen in \cite{Alestas:2021luu,Alestas:2020zol,Alestas:2020mvb} that the $H_0$ tension is alleviated due to the impact of an effective gravitational coupling evolution. As investigated in \cite{Arjona:2018jhh}, the Hu-Sawicki model studied here has an equation of state which crosses at late times from a phantom to quintessence dynamics for $b > 0$, and on the contrary for $b < 0$. Thus, we can note that the joint analysis with BAO + BBN has a best-fit with a tendency towards the quintessential behavior, i.e, $b > 0$, while for the BAO + BBN + PantheonPlus\&SH0ES combination, the best-fit 
indicates a phantom behavior. It is important to emphasize that there is no clear evidence for $f(R)$ gravity dynamics at more than 1$\sigma$. Thus, the Hubble tension is alleviated because of the effective phantom dark energy at late times induced by 
the $f(R)$ gravity dynamics.

From Table \ref{tab:all}, we notice $|\Delta \rm ln \mathcal{Z}|  < 2.5$ for all the analyses. Therefore, the Bayesian evidence of $f(R)$ models compared to $\Lambda$CDM is either weak or inconclusive. So the $f(R)$ models cannot be discriminated from $\Lambda$CDM statistically in all analysis carried out here. Finally, in Fig. \ref{mu_the_ob} we show the magnitude-redshift relation of the PantheonPlus sample for the best fit values from BAO+BBN+PantheonPlus and BAO+BBN+PantheonPlus\&SH0ES for both the $f(R)$ gravity models under consideration in the work  plus the reference $\Lambda$CDM model, where we note that  the model's predictions for low-$z$ are almost indistinguishable from each other, but it may slightly differ at high-$z$.

Finally, in Table \ref{tab:alll} we show the summary of the statistical analyses considering both the $f(R)$ gravity models investigated in this work, but without taking into account the extra correction on the distance modulus in eq. (\ref{mb_fr}), i.e, the distance modulus is now only affected due to the expansion rate of the Universe, $H(z)$, predicted by each $f(R)$ model. Thus, we can quantify statistically how much the  effective time variation of the Newton’s gravitational constant affects the observational constrains on the models. From our analyses we find that the parameter $b$ gets affected. 
For the Hu-Sawicki model, we noticed an 86\% and 18\% improvement in the constraint on $b$ when considering the effective gravitational coupling on eq. (\ref{mb_fr}) from the BAO+BBN+PantheonPlus\&SH0ES and PantheonPlus\&SH0ES joint analysis, respectively. 
Clearly a significant improvement is observed in the parameter $b$  within the Hu-Sawicki framework. 
Also, clearly the other baseline parameters are also affected by these considerations. 
While comparing the constraints between the parameters displayed in Tables \ref{tab:all} and \ref{tab:alll}, a significant and noticeable improvement on the error bars are clear when the presence of a varying effective gravitational constant is considered
on the distance modulus. Therefore, one 
can conclude that the presence of the extra correction on the distance modulus certainly improves the observational constraints on 
the cosmological parameters. 
As described above, changes on the parameter $b$  in the Starobinsky framework has low impact on the observables. Thus, for this specific case, within the formalism used in this work, this model does not show any improvements in this regard. In all cases, there was no change in conclusions referring to the Bayesian evidence calculation.

\section{Final Remarks}
\label{sec-summary}

In this work, we have considered new extra degree of freedom of the gravitational origin by modifying the gravity sector that possesses GR as a particular limit. The new scalar degree(s) of freedom are proposals of intense investigation as alternatives to $\Lambda$CDM frameworks in the last two decades. Certainly, one of the most popular theories to explain the late-time acceleration in this sense is the $f(R)$ gravity theory. In this work, we have presented an update of observational constraints with new perspectives on two well known and widely used $f(R)$ gravity models, viz., Hu-Sawicki and Starobinsky models. The robustness of the state-of-the-art assumptions on BAO+BBN data is used for the first time to constrain the dynamics of these models. Then the most recent SNe Ia data, by taking the time variation of the Newton's gravitational constant over cosmic time to correct the supernovae distance modulus relation predictions, are used in the joint analysis with BAO+BBN. Finally, the inclusion of the very low-z Cepheid host distances, including the full covariance of the SH0ES sample, is considered to investigate the $f(R)$ models under consideration in this work. We have found a minor evidence for $f(R)$ gravity under the Hu-Sawicki dynamics from BAO+BBN and BAO+BBN+uncalibrated supernovae joint analysis, but the inclusion of Cepheid host distances, makes the model compatible with GR. In general, in all the analyses, we find that  $b$ is consistent with 0 at 95\% CL for both of the $f(R)$ models. So we have not found any significant deviation from GR, i.e., $b=0$, after the application of late-time data sets. For the Hu-Sawicki model, we have noticed correlation between $b$ and $H_0$ from different observational data sets, this shows the tendency of the model to relax the $H_0$ tension. Furthermore, the free parameter $b$ of the theories still is weakly constrained which is clearly observed from its large error bars. The generalization of the perspectives considered here can be carried out with CMB data from Planck, the Atacama Cosmology Telescope and  with full-shape galaxy power spectrum sample in the light of the $H_0$ tension. The full-shape galaxy power spectrum has recently been well developed for non-standard models \cite{Philcox:2020vvt,Chudaykin:2020ghx,Nunes:2022bhn,Simon:2022adh,Carrilho:2022mon,Reeves:2022aoi,Simon:2022hpr}, including modified gravity scenarios \cite{Piga:2022mge}.
We hope to report the results in this direction in future communications.

\appendix
\section{$f(R)$ gravity in the Einstein frame}

In this appendix, we write the models under consideration in this work in the Einstein frame. In what follows, we follow \cite{DeFelice:2010aj,Sotiriou:2008rp}. Any $f(R)$ gravity metric can be written under the action

\begin{equation}
\mathcal{S} = \frac{1}{2 \kappa}\; \int d^4 x \sqrt{-g}\,\, f(R) + \mathcal{S}_{\rm m} (g_{\mu \nu}, \psi),
\label{action_2}
\end{equation}
where $\kappa = 8 \pi G$ and $\psi$ represents the matter fields. One can introduce a new field $\chi$ and write the dynamically equivalent action

\begin{equation}
\mathcal{S} = \frac{1}{2\kappa} \; \int d^4 x \sqrt{-g}[f(\chi) + f'(\chi)(R-\chi)] + 
\mathcal{S}_{\rm m} (g_{\mu \nu}, \psi).
\label{action_3}
\end{equation}

Variation with respect to $\chi$ leads to the equation $f''(\chi)(R-\chi) =0$. Thus, $\chi = R$ if $f''(\chi)=0$, which reproduces the action \ref{action_2}. Re-parameterizing the field $\chi$ by $\phi = f'(\chi)$ and setting $V(\phi) = \chi(\phi) \phi - f(\chi(\phi))$, the action takes the form

\begin{equation}
\mathcal{S} = \frac{1}{2 \kappa}\; \int d^4 x \sqrt{-g}[\phi R - V(\phi)] + 
 \mathcal{S}_{\rm m} (g_{\mu \nu}, \psi).
\label{action_4}
\end{equation}

One can perform a conformal transformation and rewrite the action above in the Einstein frame. Specifically, by performing the conformal transformation 
$\tilde{g}_{\mu \nu}= f(R) g_{\mu \nu} = \phi g_{\mu \nu}$, and the scalar field redefinition $\phi = f'(R)$ to $\tilde{\phi}$ with

\begin{equation}
d \tilde{\phi} = \sqrt{\frac{3}{2 \kappa}} \frac{d\phi}{\phi},
\label{}
\end{equation}
a scalar-tensor theory is mapped into the Einstein frame,
in which the new scalar field $\tilde{\phi}$ couples minimally to
the Ricci curvature and has canonical kinetic energy, as
described by the following action

\begin{equation}
\mathcal{S} = \int d^4 x \sqrt{-\tilde{g}} \Big[\frac{\tilde{R}}{2 \kappa} - \frac{1}{2} \partial^{\alpha} \tilde{\phi} \partial_{\alpha} \tilde{\phi} - U(\tilde{\phi}). \Big] 
\label{action_5}
\end{equation}

For the equivalent of $f(R)$ gravity metric,  we have

\begin{equation}
\phi = f'(R) = e^{\sqrt{2 \kappa/3} \tilde{\phi}}  
\label{field}
\end{equation}
and

\begin{equation}
 U(\tilde{\phi}) = \frac{R f'(R) - f(R)}{2 \kappa (f'(R))^2},
\label{potencial}
\end{equation}
where $R = R(\tilde{\phi})$. The complete and final action takes the form

\begin{align*}
\mathcal{S} &= \int d^4 x \sqrt{-\tilde{g}} \Big[\frac{\tilde{R}}{2 \kappa} - \frac{1}{2} \partial^{\alpha} \tilde{\phi} \partial_{\alpha} \tilde{\phi} - U(\tilde{\phi}) \\
& + S_M (e^{-\sqrt{2 \kappa/3}\tilde{\phi}} \tilde{g}_{\mu \nu}, \psi) \Big] 
\label{action_6}
\end{align*}

This represents a direct transformation from the Jordan frame to the Einstein frame. Now, from eq. (\ref{potencial}), we can see the type of potential for the scalar degree of freedom.
For the Hu-Sawicki model, the scalar potential is given by  \\
\begin{equation}
  U =\dfrac{\Lambda R^2}{\kappa (R+b\Lambda)^2}.\dfrac{1}{(f'(R))^2},  
\end{equation}
where $f'(R)=1-\dfrac{2\Lambda^2 b}{(R+b\Lambda)^2}$.\\
For the Starobinsky model, the scalar potential is given by  \\
\begin{equation}
  U=-\Lambda R^2(2b^2-R^2-b^2\Lambda^2).\dfrac{1}{\kappa f'(R)},  
\end{equation}
where $f'(R)=1-\dfrac{4\Lambda^3 b^2R}{(R^2+b^2\Lambda^2)^2}$.\\

Note that for $b = 0$, we get $U =$ constant and the kinetic term in the action vanishes. Then the action reduces to $\Lambda$CDM model.

\acknowledgments

 S.K. gratefully acknowledges support from the Science and Engineering Research Board (SERB), Govt. of India (File No.~CRG/2021/004658). R.C.N thanks the
CNPq for partial financial support under the project No.
304306/2022-3. S.P. acknowledges the financial support from the  Department of Science and Technology (DST), Govt. of India under the Scheme   ``Fund for Improvement of S\&T Infrastructure (FIST)'' (File No. SR/FST/MS-I/2019/41). P.Y. is supported by a Junior Research Fellowship (CSIR/UGC Ref. No. 191620128350) from the University Grants Commission (UGC), Govt. of India.

\bibliography{main}

\end{document}